# Kaula's rule – a consequence of probability laws by A. N. Kolmogorov and his school

Gledzer E. B., Golitsyn G. S.

A. M. Obukhov Institute of Atmospheric Physics, RAS

Moscow, 119017, Pyzhevsky 3, RF

e-mail: gsg@ifaran.ru

*Introduction*

At the beginning of 1930-s A. N. Kolmogorov has published three papers on analytical methods for the probability theory. The two-page work [1] had the essence of the approach started by A. Einstein and developed further by Fokker and Planck. He proposed a fundamental solution for evolution description of the 6D vector $p(u_i, x_i, t)$ of the probability distribution at the Markov character of action which in modern terms is usually called as time $\delta$-correlated acceleration. This is an approximation of processes when correlation time of random forces is much smaller than the reaction time of the system in consideration. In 1D approximation the action of such random accelerations $a(t)$ with $\langle a(t_1)a(t_2)\rangle = a^2\delta(t_1-t_2)$ can be described by the simplest equations [2]

$$\frac{du(t)}{dt} = a(t), \quad \frac{dx(t)}{dt} = u(t), \qquad (1)$$

where $u(t)$ and $x(t)$ are the velocity and displacement of a particle under acceleration.

For description of the probability $p(u_i, x_i, t)$ evolution Kolmogorov was using a Fokker-Planck type equation [1]

$$\frac{\partial p}{\partial t} + u_i \frac{\partial p}{\partial x_i} = D \frac{\partial^2 p}{\partial u_i^2}, \qquad (2)$$

where $D$ is coefficient of diffusion in the velocity space. He wrote down the fundamental solution of eq. (2) as (see also [2], §24.4):

$$p(u_i, x_i, t) = \left(\frac{\sqrt{3}}{2\pi Dt}\right)^3 \exp\left[-\left(\frac{u_i^2}{Dt} - \frac{3u_i x_i}{Dt^2} + \frac{3x_i^2}{Dt^3}\right)\right]. \qquad (3)$$

A.M. Obukhov [3] was first who analyzed and used this equation. He has shown that the coefficient $D$ is proportional to the generation/dissipation rate of the kinetic energy of the random motions $D = \varepsilon/2$. The solution (3) demonstrates that the seeked probability distribution has the normal character. To the Kolmogorov solution Obukhov added the second term at the exponent. This solution has three scales

$$\langle u_i^2\rangle = \varepsilon t, \quad \langle u_i x_i\rangle = \varepsilon t^2, \quad \langle x^2\rangle \equiv r = \varepsilon t^3, \quad \varepsilon = 2D, \qquad (4)$$

where $u^2$, $x^2$ are squares of velocity and displacement of Lagrangian fluid particles. These time dependencies have been tested numerically for ensembles of $N$ random particles [3]. It was found that even for $N = 10$ the relationships (4) started to be revealed quite satisfactorily. Moreover, the change of variables

$$t = \tilde{t}\,\tau,\ u = \tilde{u}(\varepsilon\tau)^{1/2},\ r = \tilde{r}(\varepsilon\tau^3)^{1/2}$$

transforms eq. (2) to the universal or selfsimilar form [3]

$$\frac{\partial p}{\partial \tilde{t}} + \tilde{u}\frac{\partial p}{\partial \tilde{r}} = \frac{1}{2}\frac{\partial^2 p}{\partial \tilde{u}^2}. \qquad (5)$$

The scales (4) can be interpreted as structure functions counted from zero initial conditions which allows us to introduce their spectral forms [2]. If we exclude time from the third scale (4) and put it into the first two scales we obtain the results of Kolmogorov – Obukhov of 1941 $D_u(r) = (\varepsilon r)^{2/3}$ and Richardson – Obukhov [5-7]

$$K(r) = \varepsilon^{1/3} r^{4/3}. \qquad (6)$$

These are results for the inertial interval of turbulence. The results as equations (4) are valid independent of the space dimension as has been checked by numerics [4]. It is surprising that since 1920s nobody has tried to understand why the Richardson's "law of 4/3" is observed up to 2-3 thousands km for horizontal motions, i.e. the vertical component and full velocity isotropy are not important. Lindborg [8] has shown in 1999 by treatment of experimental data of commercial flights in terms of structure functions that the dependence $r^{2/3}$ is traced up to distances 2-3 thousands km. His results in terms of the eddy diffusion explain the Richardson results of 1920-s (see [8]). All can be explained starting with a two-page note of Kolmogorov, 1934 [1], and subsecond development of his results by Obukhov [3] and Monin and Yaglom [2]. Only the Markov properties of the probability distribution is important, not the full isotropy. Our numerics of 2010 [4] has confirmed this independence on the space dimensionality for the second moment with the accuracy of the numerical coefficient up to O(1).

For further development we need an analytical connection between these second moments, and their spectral representations when both of them are power laws. Consider first the structure function $D(t) = At^\gamma$, $0 < \gamma < 2$. The relation to spectrum is [2]

$$D(t) = 2\int_0^\infty (1 - \cos \omega t) E(\omega) d\omega, \tag{7}$$

which is

$$E(\omega) = \frac{C}{\omega^{\gamma+1}}, \quad C = \frac{A}{\pi}\Gamma(\gamma+1)\sin\frac{\pi\gamma}{2}, \tag{8}$$

where $\Gamma(\gamma+1)=\gamma\Gamma(\gamma)$ is the gamma function. The length scale is related to the mean square displacement of a particle has therefore the spectrum $C\omega^{-4}$, $C = A/\pi$ and for the velocity scale $\langle u^2 \rangle = \varepsilon t$ the spectrum is $\varepsilon\omega^{-2}$. According to the terminology introduced by Yaglom [9, 10] the spectrum $\omega^{-2}$ corresponds to a process with stationary increments of the first order ($0 < \gamma < 2$), and the spectrums $\omega^{-4}$ has the second order stationary increments ($2 < \gamma < 4$). In order to avoid singularity at $\omega \to 0$ the transformation kern in eq. (6) is $(1 - \cos\omega t)^n$, $n = 1$ for the velocities and $n = 2$ for coordinates.

*Description of the relief's statistical structure*

When a flying altimeters measures the relief we obtain a temporary signal $h(t)$. The time $t$ here is related to horizontal coordinate by $y = ut$, $u$ being the altimeter velocity. The standart Fokker – Planck equation with stochastic velocity fluctuations is in this case

$$\frac{\partial p}{\partial t} = D\frac{\partial^2 p}{\partial h^2}$$

with $y = ut$ it is transformed into

$$\frac{\partial p}{\partial y} = D_1 \frac{\partial^2 p}{\partial h^2}, \quad D_1 = \frac{D}{u} \tag{9}$$

The diffusion coefficient $D_1$ has the dimension of length. But if we use different dimensions for vertical $L_z$ and horizontal $L_y$ then $D_1 = L_z^2/L_y$ revealing clearly the diffusion character of the relief formation process. The eq. (9) has the scale, the structure function with zero's initial condition

$$h^2 = D_1 y$$

with the corresponding spectrum

$$S(k) = D_1 k^{-2}, \quad k = 2\pi/\lambda y \tag{10}$$

4$\lambda$ being the corresponding wave length which is applicable for small scale continuous areas. The spectrum of the relief slopes in this approximation $S_\zeta(k) = k^2 S(k) = D = $ const. The constancy of a spectrum is called "white noise". The slope of the relief

$$\frac{dh(y)}{dy} = \varsigma(y) \qquad (11)$$

is stochastic slope for which we accept the hypothesis of $\delta$-correlation with horizontal coordinate:

$$\langle \varsigma(y_1)\varsigma(y_2) \rangle = \varsigma^2 \delta(y_1 - y_2) \qquad (12)$$

which corresponds to the Markov character of forces in [1-3]. Here $\zeta^2$ may be interpreted as the mean square slope angle, a nondimensional value. Generally speaking the right side of eq. (11) should be $\frac{1}{2} r_y \varsigma^2$, where $r_y$ is the horizontal correlation slope scale, but we assume that it much smaller then relief large scale variations.

Now we return to concrete planetary reliefs. Recently an extensive paper on ultrahigh spherical harmohics analysis for topography of Earth, Mars and Moon has appeared [11] with horizontal resolution up to hundred of meters. Much earlier analyses of topography for Earth and Venus are in [12]. In this book there are data for mountain, hilly and plain areas for hundred kilometers in size at the Oregon state. The flights were in various directions in each area. For records in the range 1 – 60 km the mean value of these spectra were found [13] to be 2.03 ± 0.04. Unfortunately there were no data in [12] for spectral amplitudes.

In the book [12] the spectral index 2 (and other statistical measures, as Hausdorff measure, the fractal index $d$) are found by original records treatment, but no analyses why they are of particular concrete values. The analyses by Kolmogorov [1-2] had been proposed by several decades earlier and are clearer and simpler. As the base the only $\delta$-correlation is ised for processes in consideration, i.e. Markov's hypothesis. Such an assumption is used in many branches of theoretical physics meaning only that the system reaction time is much larger than the correlation time of stochastic influence on it.

*Spherical harmonics*

Consider the relief on a sphere $z(\varphi,\theta)$ with a radius $r$, $\varphi$ is longitude, $0 \leq \varphi < 2\pi$, $\theta$ is a co-latitude, $0 \leq \theta \leq \pi$. Then

$$\frac{\partial z(\varphi,\theta)}{r\partial \theta} = \varsigma_\theta(\varphi,\theta)$$

is the slope angle. Introduce $x = \cos\theta$ and present the relief by normalized associated Legendre polinomials $P_j^{|m|}(x)$

$$\frac{z(\varphi,\theta)}{r} = -\sum_{j=1}^{\infty}\sum_{m=-j}^{j} a_m^j \exp(im\varphi)\Phi_j^{|m|}(x), \qquad \Phi_j^{|m|}(x) = N_j^{|m|} P_j^{|m|}(x), \qquad (13)$$

$$\int_{-1}^{1}\Phi_j^{|m|}(x)\Phi_i^{|m|}(x)dx = \delta_{ji}, \qquad N_j^{|m|} = \left(\frac{2j+1}{2}\frac{(j-|m|)!}{(j+|m|)!}\right)^{1/2}.$$

Then due to eq. (12) the relief slope has to be

$$\varsigma_\theta(\phi,\theta) = \sum_{j=1}^{\infty}\sum_{m=-j}^{j} a_m^j \exp(im\varphi)\sin\theta \frac{d\Phi_j^{|m|}(x)}{dx}. \qquad (14)$$

Consider the mean normalized energy of the relief slopes:

$$E_\theta = \int_0^{2\pi}\int_{-1}^{1} <\varsigma_\theta^2> d\varphi dx = 2\pi\sum_{j=1}^{\infty}\sum_{m=-j}^{j} <|a_m^j|^2> \int_{-1}^{1}(1-x^2)\left(\frac{d\Phi_j^{|m|}(x)}{dx}\right)^2 dx. \qquad (15)$$

Mean values of the spectral components $<|a_m^j|^2>$ we consider to be not depending on $m$, which is confirmed by observations and on practice [12] is presented only on index $j$, and are equal to

$$<|a_m^j|^2> = \frac{\alpha^2}{j(j+1)(j+1/2)}, \qquad (16)$$

where $\alpha$ is the mean angle of the relief slope. Accounting for this after calculation integrals in (15) and summing up over $m$ we have from (15)

$$E_\theta = 2\pi\sum_{j=1}^{\infty} S_j, \qquad S_j = \alpha^2, \qquad (17)$$

or constant spectrum of "white noise", i.e. $\delta$-correlation with $\theta$. Then for the relief energy from (12)

$$E = \int_0^{2\pi}\int_{-1}^{1} <z^2> d\varphi dx = 2\pi r^2 \sum_{j=1}^{\infty}\sum_{m=-j}^{j} <|a_m^j|^2> \int_{-1}^{1}(\Phi_j^{|m|}(x))^2 dx = 2\pi r^2 \sum_{j=1}^{\infty}\sum_{m=-j}^{j} <|a_m^j|^2>, \quad (18)$$

wherefrom it is clear that the mean of the spectral components (16) gives at $j \gg 1$ for the relief spectrum $\sim j^{-2}$, as it was found from empirical data [12, 14-17]

$$E = 2\pi \sum_{j=1}^{\infty}\sum_{m=-j}^{j} \frac{r^2 a^2}{j(j+1)(j+1/2)} = 2\pi \sum_{j=1}^{\infty} \frac{r^2 a^2 (2j+1)}{j(j+1)(j+1/2)} = 4\pi \sum_{j=1}^{\infty} \frac{r^2 a^2}{j(j+1)} = 4\pi r^2 \alpha^2 = 4\pi r D_1, \quad (19)$$

where $D_1$ is the horizontal diffusion coefficient and also has been note that $\sum_{j=1}^{\infty}\frac{1}{j(j+1)} = 1$ (use $\frac{1}{j(j+1)} = \frac{1}{j} - \frac{1}{j+1}$ identity). In order to come from a natural number index $j$ to continuous wave number $j$ at $j \to \infty$ we shall use in eq. (13) an asymptote of the polinome $P_j^{|m|}(\cos\theta)$ at $j \to \infty$:

$$P_j^{|m|}(\cos\theta) \approx (-1)^m \left(\frac{2}{\pi j \cos\theta}\right)^{1/2} \cos\left(\alpha - \frac{\pi}{4} + \frac{m\pi}{2}\right), \quad \alpha = \left(j+\frac{1}{2}\right)\theta \quad (20)$$

Now instead of co-latitude $\theta$ we will use the length $y$ along meridian starting at north pole:

$$y = r\theta, \ 0 \le \theta < \pi. \quad (21)$$

Now we rewrite the phase $\alpha$ in (20) as $\alpha = y(jdk + dk/2)$, $dk = \frac{1}{r}$ is an increment of the wave number $k = j \cdot dk$ at $j \to \infty$. Then at $dk \to \infty$:

$$\alpha = ky, \ k = = j \cdot dk = j/r \quad (22)$$

Then use the wave number $k$ in the eq. (18) for the relief energy and at $j \to \infty$ obtain

$$E = 2\pi D_1 k^{-2} = 4\pi r \alpha^2 k^{-2}.$$

Here $4\pi r \alpha^2 k^{-2} = S(k)$ is the spectrum with dimension of the cubic length (recall that $\alpha$ the mean slope angle). It can be seen that the value $r\alpha^2$ in (23) is proportional to the diffusion coefficient $D_1$ in the coordinate space.

The relief spectrum in eq. (19) differs from the Kaula's "rule of the thumb" $\propto j^{-2}$ by the multiplyer $[j(j + 1)]^{-1}$. The relative difference between the two at $j \to \infty$ decreases as $[j^2(j + 1)]^{-1}$ but for small $j$ it is quite noticeable, e.q. at $j = 1$ the difference is twice, and at $j = 4$ it is 20%. This shows that the hypothesis on the "white noise" of the relief slope angles is not acting and the corresponding relief at structure at $j \leq 5$ could be different at each celestial body due its peculiar tectonics.

*Discussion of the results*

Our scheme of the relief evolution basing at eq. (8) of the probability density changes is the horisontal diffusion of the tectonically formed vertical structure under the gravity acting along slopes which resist to winds, water and rocks are running down, etc. Fig. 1 (Fig. 7.19 from [12]) presents reliefs spherical harmonics: for Earth $j \leq 180$ and Venus $j \leq 50$. With $j \geq 4$ all coefficients approach well to inverse square dependence. The scattering is noticeable especially if one recalls that data are logariphmical.

Nevertheless we compare our theoretical expressions for harmonic coefficients from eq. (19)

$$S_j = \frac{4\pi r D}{j(j+1)}$$

with their empirical values at Fig. 1.

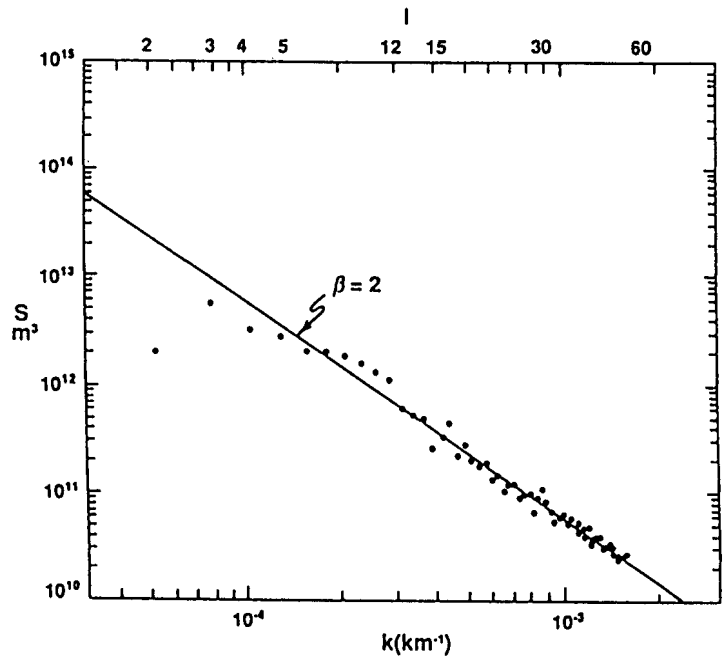

Fig. 1. Relief spectral harmonics for Earth and Venus after [12].

The comparison has been done for selected harmonics with accuracy about 10% within their values order. The harmonic number for Earth were 5, 10, 20, 30, 40, 60, 90. This has produced $D = 1.3 \pm 0.3$ m. For Venus the numbers 5, 10, 15, 20, 30, 40 and 50 producing $D = 0.19 \pm 0.03$ m. The accuracy in determination of our diffusion coefficient was about 20%, but the striking difference between the two planets has been already noted in [12]. However one should recall that Venusian data are related only to equatorial plates [12]. The global relief characteristics can be estimated with much larger precision and effort (see e.g. [11]) which we leave for much younger colleagues, but we believe that it would very much worth to do.

Fig. 2 presents the relief spectra as functions of the surface linear wave length.

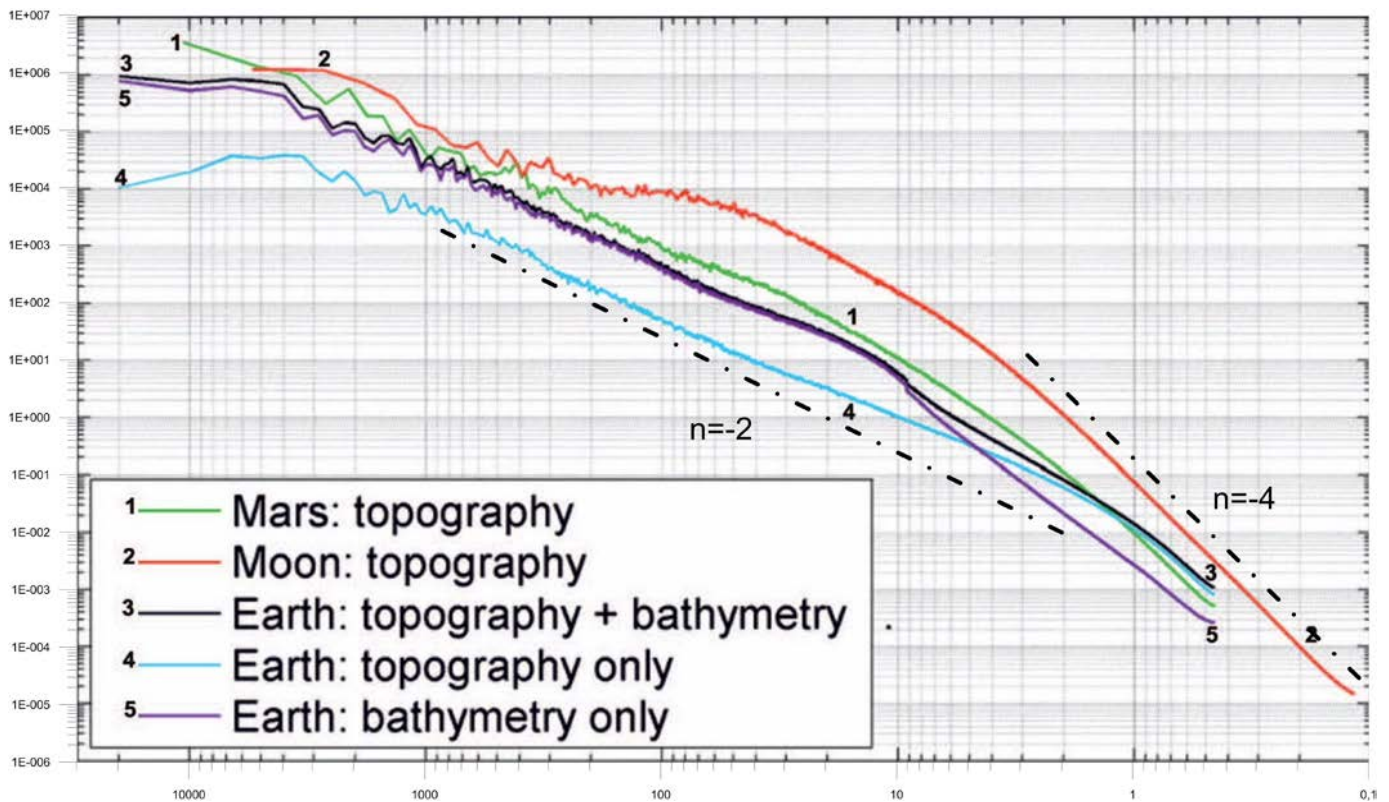

Fig. 2. Relief spectra as functions of the surface wavelength for Moon, Mars and Earth (3 cases labeled at lower left). We added lines $k^{-2}$ and $k^{-4}$ for large and small distances.

The authors of [11] have computed spherical harmonics for Earth and Moon up to $j = 46200$ and for Mars to $j = 23100$. Our eqs. (20) – (23) present high frequency harmonics transforming into $k^{-2}$. At Fig. 2 we added direct lines corresponding to spectra $k^{-2}$ and $k^{-4}$ for very high wave numbers. The dependence $k^{-2}$ is satisfactorily for Mars, line 1, and for three terrestrial spectra, lines 3-5, with some steeping for waves lesser than few kilometers. The Moon spectrum is specific at 200 – 80 km where its spectrum slope is decreasing, almost flattening. This is evident tracers of Moon bombardement by asteroids of sizes up to 10 km, because then crater sizes are 10-20 times larger than the diameter of bombarding body. Minimal lunar wave length is 120 m and down to this size starting with 3 km the spectral slope is close to –4. For other 3 spectra at Fig. 2 the slope increase is also clear. These effects can be understood if we omit the basic hypothese on the $\delta$-correlation of slopes along the horizontal at smaller scales. E.g. if we would use the high frequency spherical harmonic asymptote (20) and use instead of constant angle $\alpha$ a correlation function for angles, e.g. linearly decreasing down to a scale $\Delta$, then we could obtain an asymptote $k^{-4}\Delta^2$.

*Summing up the performed*

Our analysis is based on the Fokker – Planck – Kolmogorov, FPK, equation transformed into eq. (8) for the altimeter record. It has the solution $\langle h^2 \rangle = Dy$ for the mean square of the relief hight: where $D$ is the horizontal diffusion of the relief. Such a solution has the spectrum at $S(k) = Dk^{-2}$, $k = 2\pi/\lambda$. The modified FPK equation is based here on the hypothesis of the time $\delta$-correlation of the acting forces which in this case of relief is horizontal $\delta$-correlation of relief slope angles. The spherical harmonics of the relief have been found by Kaula to follow the inverse square rule for the harmonic number $j$. Our hypothesis on the "white noise" horizontal spectrum for the slopes confirms the Kaula rule for high harmonics but modifies slightly it to the $[j(j + 1)]^{-1}$. This difference is decreased fact as $[j^2(j + 1)]^{-1}$.

Ruther simple estimates of diffusion coefficients $D$ have been performed for Earth and Venus, the first one was found to be equal to 1.3 m, several times higher than for the second planet. A similar analysis may be performed also for planetary gravity field fluctuations with the same results though in that case one should consider also internal tectonics and/or surface processes for smaller celestial bodies under the action of environment. The consideration of statistical scales (4) as structure functions with zeroth initial conditions allows one to propose spectral representation for them.

One may conclude that the Kaula's rule for gravity field as well for the relief is an asymptotic consequence of the random walk laws, first formulated by Einstein and in the most general and practical way by Kolmogorov in 1934.

Geophysical interpretation of these results is waiting for a thorough analysis. As the Kaula's constant for gravity field is used to obtain information on the internal structure of celectial bodies (mascones, isostasy), the diffusion coefficient $D$ could contain information on tectonics, surface material properties, etc. This coefficient is an analogue for the Kaula's gravity field fluctuations constant. Much work has to be done by much younger people because the total age of these two authors is above 157 years.

This work has been particularly supported by the RAS Presidium Program №7 "The development of non-linear methods in theoretical and mathematical physics". The authors are very thankful to Dr. O. G. Chkhetiany for discussion of the results and his help during long work.